\documentclass[12pt,showkeys,a4paper]{article}
\usepackage{amsmath}
\usepackage{amssymb}
\begin{document}

\begin{center}
\vspace{0.5 cm} {\huge {The Inflaton and its Mass}}

\vspace{1 cm}

{\large {R. Brout}}\\ Physique th\'eorique U.L.B. CP 225\\ Bvd du Triomphe\\ 1050
Bruxelles Belgium\\
\vspace{1cm}

\end{center}
\vspace{0.5cm}

\begin{abstract} In the context of the two fluid model of space-time fluctuations
proposed to tame the transplanckian problem encountered in black hole physics, it
is postulated that the inflaton is the fluctuation of mode density, ``the vapor
component'' of the model. The mass of the inflaton is occasioned by the exchange of
degrees of freedom between the ``vapor'' and the ``liquid'', the planckian ``soup''
in which usual ``cisplanckian'' fields propagate. This exchange between vacuum
fluctuations is modeled after its counterpart in the real world i.e. black hole
evaporation. In order of magnitude, a very rough semiquantitative estimate, would
situate the mass somewhere between $10^{-10}$ and $10^{-5}$ planck masses, the
largest uncertainty being the mass of the planckian black hole fluctuation i.e. the
entropy that one ascribes to it.
\end{abstract}
\newpage
\par In ref. \cite{LABEL1}, it was proposed that the inflaton field in the scenario of
chaotic inflaton could be interpreted as the fluctuation of the density of modes of
the quantum fields used to describe matter in vacuum (zero temperature).  These are
``cisplanckian'' modes, cut off at a planckian mass scale. The idea is based on the 2-fluid model,
a ``vapor of modes'', and a planckian substratum, a ``soup'' in which the modes propagate. This
model has been called upon, in one form or another, to tame the so-called transplanckian problem
that arises in the theory of black hole evaporation which appears in the backward extrapolation of
modes that give rise to Hawking evaporation. These modes crowd into a very narrow region around the
horizon resulting in an exponential growth of their proper energy, far exceeding the planck mass.
See ref.
\cite{LABEL3} for an overall view of various efforts to confront this situation.
\par In this paper we shall analyze in more detail than in ref. \cite{LABEL1}, the dynamics
of the inflaton that emerges in this view. Though the present
considerations were concieved independently, they should be considered as a complement to some
recent cosmological applications of the 2-fluid model\cite{LABEL031}. In these latter works, the
problem was addressed as to how modes are produced as the universe expands to keep their
{\underline{mean}} density fixed. Since this mean arises from dynamics it is necessarily 
acccompanied
by fluctuations; ergo our inflaton! The principal question is how to explain
the inflaton mass which in phenomenological analyses is ($10^{-5}-10^{-7})m_{Pl}$. Since our
knowledge of the planckian substratum is so limited, we shall pursue a
phenomenological approach based on the exchange of degrees of freedom between the
two fluids. This mechanism is in harmony with the ideas of refs \cite{LABEL3,LABEL4,LABEL5} wherein
a mode that approaches the horizon dissipates into the ``soup''. It is shown that a
reasonable set of hypotheses results in an acceptable order of magnitude for the
inflaton mass.

\par Of course the central problem of quantum gravity is to find the ``theory of
soup''. The most developed ideas in present theory, though still quite far from the
ultimate goal, occur in string theory involving dynamically engendered structures
in higher compactified dimensions. The dissipation might then be the dissipation
of the modes into higher dimensions. As an explicit example, one may conjecture that this is
encoded in the deformed commutator of ref.\cite{LABEL4}. This resulted in an apparent violation
of unitarity in the description  of mode propagation near the black hole horizon. It is also
appropriate to point out here that it is currently thought that the Hagedorn phenomenon encountered
at energy scales near the string tension causes the strings to fold up into planckian black holes
\cite{LABEL6}.
\par Other efforts to model soup are through the use of gravitational instantons
\cite{LABEL7} or black hole pair fluctuations \cite{LABEL8}. Whatever, the strength of gravitational
interaction at the planckian level renders a description, in terms of modes,
inappropriate. Rather, it would seem that a planckian soup, whose structural
elements are of black hole character, is closer to the truth. In what follows we
shall adhere to this hypothesis.
\par One may well question the usefulness of indulging in theoretical speculation
concerning physical problems which is based on such ignorance.
`` To what purpose disturb the dust on a bowl of rose leaves?\footnote{T.S.
Eliot ``Burnt Norton''}''. To which one may respond : if the answer to a physical
question depends partially on the unknown, then to elaborate hypotheses concerning
its incidence on the observable world is a legitimate logical exercise which allows
one to explore these hypotheses as a guide towards a more fundamental theory.
\par Armed thus with the experience of the 2-fluid approach applied to  black hole
evaporation, it is then natural to inquire into what it can offer for cosmogenesis
and inflation. It is in this spirit that we explore the possibility of identifying
the inflaton with mode density.
\par In no small measure this effort stems from our desire to formulate
cosmogenesis principally in terms of gravity rather than the symmetries (better,
broken symmetries) of particle physics. The scenario of chaotic inflation
\cite{LABEL91} has done much to point the way. A scalar field, $\phi$, has a (rare)
fluctuation in a patch of space-time, of sufficiently large amplitude and extension
so as to be seized upon by the cosmological component of gravity, the scale factor,
and to inflate. So this construction has the pleasing epistemological conception of
ascribing to our universe a beginning. Moreover, quite naturally the inflation, so
induced, has an end. Regression of the fluctuation gives way to ``cisplanckian''
particle production and the subsequent adiabatic era.
\par All well and good! But who is the inflaton $\phi$? And how can one come by
such an unlikely mass as $10^{-7}$ to $10^{-5} m_{Pl}$ for his (or her) mass, this
being the value one comes upon when applying the chaotic inflationary scenario to
model the presently observed CMBR fluctuations and galaxy distributions? This
latter question is most disconcerting. The number $10^{-5}$ is neither hither 
nor yon. Gravity is on the scale of 1. GUTS requires a much higher mass than
$10^{-5}$ elsewise the proton would have been seen to decay a long time ago (unless
the usual formulations are contorted unnaturally to give a selection rule against
proton decay). And supersymmetry is supposed to operate on the $Tev$ level (i.e.
$10^{-16}$) in order to come to grips with the hierarchy problem of particle
physics. It would seem that nature has no place for $10^{-5}$.
\par Often in physics when such an unlikely number appears, it corresponds to an
unlikely event, or better put an improbable event, such as tunneling.  Small numbers
are exponentials of numbers of $O(1)$. An example related to tunneling is black
hole evaporation. (One method of computation involves a steepest descent integration
with a saddle point existing at a complex time (10)). This consideration will play
an important r\^ole in our effort to account for the inflaton mass.
\par Let us now begin to formalize the 2 fluid hypothesis. The energy density of
the fluid is written 
\begin{equation}
\label{1.1}
\epsilon = \epsilon_M + \epsilon_P + \epsilon_{MP}
\end{equation} the sum of mode energy $(\epsilon_M)$, soup energy $(\epsilon_P)$ and their
interaction $(\epsilon_{MP})$. For
$\epsilon_M$ one estimates $\epsilon_M = O(\Lambda^4)$, $\Lambda$ being a planckian
cut-off. One expects $\epsilon_P$ to be the same order of magnitude, but in view of
our previous discussion, it is at present difficult (to say the least) to come by.
It is $\epsilon_{MP}$ which will play an important role in our attempt to estimate
the inflaton mass. Indeed one of the important contributions to the fluctuation of
mode density is expected to be the passage of degrees of freedom between the two
fluids and it is this that engenders the mass in our scheme. First, however, let us
model the fluctuations in the ``fluid of modes'' alone without exchange with the
planckian reservoir.
\par Since this fluid of modes is conceived to be a dynamical entity in itself it
is subject to density fluctuations. Lest there be any misunderstanding, this
fluctuation is not that  of a mode, such as a phonon in a fluid or solid or a
quantum in a quantum field theory. In that case the zero point energy (i.e. the
mean energy of modes in vacuum) is fixed by the cut-off. We are now allowing the
cut-off to fluctuate. This occurs because we consider the density of the underlying
soup to fluctuate and the cut-off occurs when the momentum of the mode is
$(O(n^{1/3}_P)$ where $n_P$ is the density of soup. One of our main postulates
is that this whole complex is in (meta) stable equilibrium. The reason for (meta)
is that in cosmology a fluctuation sufficiently far from equilibrium will inflate
before it regresses. In what follows we abstract and ignore this, thereby
characterizing only small fluctuations, those that occur in conventional fluid
dynamics near equilibrium.
\par This leads to a review of the conventional theory of sound, albeit in terms
suitable for our purpose.
\par Let $<n>(\equiv n_0)$ be the mean density of modes, $O(\Lambda^3)$, and 
$\delta n$ its fluctuation. In view of our variational ansatz, the potential energy
engendered by $\delta n$, up to quadratic terms is 
\begin{eqnarray}
\label{1.2} V&=& {1\over 2} \int d^3 r[\left .\partial^2\epsilon/\partial
n^2\right|_{n=n_0}(\delta n)^2]\\ &\equiv& {1\over 2} B \int (\delta n)^2\/d^3r
\end{eqnarray} One thinks of $B$ as a bulk modulus since the pressure fluctuation
induced by $\delta n$ is
\begin{equation}
\label{1.3}
\delta p  = n_0 \delta(\partial \epsilon/\partial n) +\cdots = (Bn_0) \delta n+\cdots
\end{equation} (since $<p>=\partial \epsilon/\partial n|_{n_0} = 0$ by our ansatz).
\par When taken together with the kinetic energy, $T$, the theory of sound then
follows. A convenient way to proceed is through the displacement field, $\vec
\chi$ which is related to $\delta n$ through the equation of continuity
\begin{equation}
\label{1.4}
\delta n = -n_0 \vec \nabla.\vec \chi
\end{equation} (More familiarly, in terms of $\vec j (= \vec{\dot \chi})$ one has 
$\delta \dot n = -n_0 \vec \nabla.\vec j$ to first order in the fluctuation).
\par $\vec \chi$ (or $\vec j$) has been introduced since in terms of it the kinetic
energy, $T$, is the simple form 
\begin{equation}
\label{1.5} T = \frac{\rho_0}2 \int d^3r(\partial \vec \chi/\partial t)^2
\end{equation} where $\rho_0$ is the mass density at equilibrium. From Eqs
(\ref{1.5}) and (\ref{1.2}), and relating $\delta n$ and $\chi$ through (\ref{1.4})
leads to 
\begin{equation}
\label{1.6} {\partial^2\vec \chi\over{\partial t^2}} - c^2 \vec \nabla (\vec \nabla
.\vec \chi)=0
\end{equation} or from (\ref{1.4})
\begin{equation}
\label{1.7} {\partial^2\delta n\over{\partial t^2}} - c^2 \vec \nabla^2 \delta n = 0
\end{equation} where
\begin{equation}
\label{1.8}
\rho_0c^2 = n^2_0 \partial^2\epsilon/\partial n^2
\end{equation}
\par [Surprisingly, at least to the author, $T$ is not a local quadratic  form in
$(\partial \delta n/\partial t)$. Rather it is non local $=(\rho_0/2)
\int \delta \dot n (\nabla^2)^{-1} \delta \dot n$. ]
\par We adapt this non relativistic analysis to the relativistic situation wherein
every massless field propagates with a common velocity, $c$, which, as usual, we
set to unity. Then $\rho_0$ is constrained by covariance to be Eq. (\ref{1.8})
with $c=1$. The canonical field is $\rho^{1/2}_0
\vec \chi$.
\par That $\vec \chi $ (or $\delta n$) is massless in the conserved case is evident
since there is then no fluctuation without compensating gradients i.e. in Fourier
space $\delta n(k) = \omega (k) = 0$ as $k\rightarrow 0$, where $k$ and $\omega$
are wave number and frequency. Mass is induced by source terms in the equation of
continuity, induced by exchange with the reservoir. Thus 
\begin{equation}
\label{1.9}
\delta n = -n_0 \vec \nabla.\vec \chi+\delta n_s
\end{equation} where $\delta n_s$, is the source.
\par We must then address the question of assessing the modification that
$\delta n_s$ will induce on the wave equation (\ref{1.7}). Without recourse to a
reliable model of the ``planckian soup'' and the exchange mechanism with the
``vapor of modes'' one must proceed phenomenologically making use of a reasonable
set of postulates.
\par As in the massless (i.e. conserved) case we study small fluctuations. In so
doing we maintain the linearity of the equation that describes the evolution of the
fluctuation. This may not be sufficient to describe the fluctuation to which  one
appeals in setting up chaotic inflation, but it will allow for a first estimate of
the order of magnitude of the inflaton mass. And then one can hope to proceed to
stronger fluctuations at a later stage.
\par The second ingredient of general character that we shall use is covariance.
Again, this may not reflect the whole truth of the matter. Cosmology is carried out
within a rather rigid frame. This is given by the nested set of surfaces of
homogeneity related one to the other by the expanding scale factor. And in the
hamiltonian formation this latter is proportional to time itself. All theoretical
developments of cosmology are effectively welded into this frame. It is therefore
not obvious that covariance is a sure guide when one considers the primordial
fluctuation which has been postulated to be at the origin of this highly
constrained theater of operations. In what follows we shall nevertheless postulate
covariance since, once more, it paves the road towards an estimate of what we are
aiming for, the inflaton mass. In the last paragraph of this paper we shall return
briefly to the question of what other physical effect might come into play when
one lets up on the principle of covariance.
\par From these two postulates, it follows that one should add a term proportional
to $\delta n_s$ to Eq. (\ref{1.7}) (with $c^2=1$), since $\delta n_s$ is envisioned
to be scalar. Moreover, it should not contain terms like $ \square \/\delta n_s$.
This is because we are seeking an effect at small, or even vanishing $k$. More
precisely, we envision extended fluctuations in space whose profile could be
nodeless due to the exchange of degrees of freedom over this extended region. Such
a fluctuation has an important Fourier component at $k=0$. It will therefore cause
this component to oscillate, the  hall-mark of a mass term. To find this latter, we
must set up a relation between $\delta n_s$ and $\delta n$ itself.
\par From the above and the assumed linearity we may then write
\begin{equation}
\label{1.10}
\delta \ddot n (k=0) = -\mu^2 \delta n (k=0)
\end{equation} where the r.h.s. appears as a consequence of adding a term in
$\delta n_s$ to the wave equation. The mass, $\mu$, is positive or negative
according to whether the degree of freedom comes (or goes) from (into) the
reservoir. Since we are not considering derivatives of $\delta n_s$, the
implication is $\delta n_s$ and $\delta n$ are proportional. A nice physical
understanding of this proposition follows from an interpretation in second
quantization.
\par The equation which is being proposed is 
\begin{equation}
\label{1.11} [\square + \mu^2] \delta n = 0
\end{equation} where the $\mu^2$ term arises from the (linear) addition of $\delta
n_s$. To facilitate the discussion in terms of the modes of $\delta n$ let us
introduce a large arbitrary volume $V$, for purposes of normalization. [ Lest there
be any confusion, ``mode'' in the present context means eigensolutions of
(\ref{1.11}) not the original quantum modes which go to make up the ``vapor of
modes'']. Each mode $\delta n(k)$ of (\ref{1.11}) obeys an oscillator equation,
hence has non vanishing matrix elements (in $V$) between $N(k)$ and $N(k)\pm 1$ 
where $N(k)$ are integers. $\delta n_s$ and $\delta n$ are proportional if they
have the same matrix elements, up to a constant. Our assumption of proportionality
thus has the physical content that exchange with the source term occurs one by one.
\par So much for the kinematical discussion of why and how the inflaton mass can
arise in our picture. We must now try to model $
\mu$ to estimate how much.
\par At first sight one is tempted to identify $\mu$ with the energy of the degree
of freedom that is exchanged. However this cannot be the whole story. One need but
imagine a normal liquid which is externally perturbed by a source which creates or
annihilates particles. Clearly if these events are rare on the scale of wavelength
and frequencies under study, the propagation will not be much affected. It is
necessary to take into account  the probability of these events.
\par Thus we are led to postulate that $\mu^2$ is the (mass)$^2$ of the exchanged
degree of freedom multiplied by the probability that the exchange, in fact, takes
place and we write
\begin{equation}
\label{1.12}
\mu^2\tilde = p m^2_{Pl}
\end{equation} (since the mass scale of everything in sight is $O(m_{Pl})$).
\par [In ref. \cite{LABEL1}, the (mass)$^2$ of the inflaton was erroneously
identified with $[\partial^2\epsilon/\partial n^2|_{n=n_0}]$. However it should now be
clear that this potential energy term issues from both acoustic pressure and
exchange. It is only when the two effects are separated that one begins to acquire
some understanding of the origin of the inflaton mass.]
\par Let us go somewhat more deeply into the physics behind Eq. (\ref{1.12}), once
more through a quantum mechanical interpretation. Suppose Eq. (\ref{1.10}) arises
from the Fourier decomposition of an extended nodeless fluctuation. Extended means
over a scale that contains a statistically large number of planckian cells, hence of planckian
entities that make up the soup. Though the energy that each of these latter gives up to the
``vapor'' phase (or takes away) is $O(m_{Pl})$, this exchange must be weighted with
the probability that its quantum configuration necessary to effect the exchange is
realized, hence the factor $p$ of Eq. (\ref{1.12}).
\par What is $p$? In the Clausius-Clapeyron relation for the vapor pressure over a
liquid it is $\exp[-\Delta S]$ where $\Delta S =$ entropy difference per molecule
between liquid and vapor [see for example ref. \cite{LABEL10} p. 45-6]. Inspired thus, we now take
seriously a planckian substratum which, in some form or other, is modelled as a
collection of black holes. Whatever this may turn out to be in the ultimate theory
we (or at least the author) would be surprised if the entities which comprise the
soup do not have horizons, a locus which precludes classical communication between
exterior and interior. We are thus led to postulate that these entities have
entropy, and, for want of something better, we take this to be the Hawking black
hole entropy (= horizon area/4). See ref.~\cite{LABEL11} for a discussion of this
formula in a more general context than usually presented. It is also to be recalled
that, in the one example where the black hole entropy could be explicitly
calculated in string theory in terms of number of states, one again recovers the
Hawking area law \cite{LABEL12}.
\par We are thus postulating that vacuum fluctuations on the planckian scale are
black holes and that they are endowed with an entropy, in the same way as real
on-mass shell black holes.
\par In this same vein, we postulate an exchange between these planckian black hole
fluctuations and cisplanckian fluctuations, which is the analogue of black hole
evaporation, i.e. the exchange of degrees of freedom between a real black hole and
the quanta of modes. A black hole fluctuation is considered to be a dynamical
entity, a gravitational-matter field complex which can sometimes manifest itself on
the low momentum scale as having a virtual cloud of cisplanckian modes in its
surroundings. As in usual quantum mechanics this is described by matrix elements in
Hilbert space, interpreted as virtual processes. It is postulated that the
probability for this to happen (or if one wishes, the weight of the configuration
of black hole + cisplanckian mode) is the same as that encountered in the theory of
black hole evaporation $(=e^{-\Delta S})$. We are thus led to a Clausius-Clapeyron
type relation as might have been suggested at the outset from our two fluid model. 
\par Thus we write
\begin{equation}
\label{1.13}
\mu^2 = e^{-|\Delta S|}m^2_{Pl}
\end{equation} where $\Delta S$ is the difference in black hole entropy occasioned
by the evaporation of a quantum, here identified with the frequency of a typical
mode, something less than but $O(\Lambda)$.
\par The reason for absolute value of $\Delta S$ in Eq. (\ref{1.13}) is that there
is equal probability for exchange in the two directions, owing to equilibrium
conditions that govern the system about which the fluctuation takes place.
\par Eq. (\ref{1.13}) then is the explanation within the 2 fluid model of why
$\mu< m_{Pl}$. If one is so bold to put in numbers, then, with $M$ = typical mass
of a black hole fluctuation and $\bar \omega$ = typical mode frequency, one has
\begin{eqnarray}
\label{1.14}
\mu &\simeq& \exp [-4\pi M\bar \omega/m^2_{Pl}]m_{Pl}\\ &\simeq&
10^{-[5(M\bar\omega/m^2_{Pl})]} m_{Pl}
\end{eqnarray} This formula is presumably sensible only for $M>m_{Pl}$. Since
$\bar \omega = O(\Lambda) = O(m_{Pl})$, we would expect $\mu$  to range between
($10^{-5}$ and $10^{-10}m_{Pl}$ as a rough guess.
\par Though this quantitative estimate is to be taken with a grain of
salt, an optimist will seek encouragement from it. Clearly it is now necessary to
plumb detailed models to the extent that this is possible.
\par For the convenience of the reader we list below our hypotheses
\begin{enumerate}
\item Chaotic inflation implemented by the inflaton field
\item The inflaton field identified with mode density in the context of a 2-fluid
model.
\item The mass of the inflaton originating in the exchange of degrees of freedom
between the planckian substratum and the fluid of modes. The equation of
propagation for the inflaton follows from linearity (small fluctuations) and
covariance.
\item The exchange is governed by processes in vacuum which mirror black hole
evaporation in the real world.
\end{enumerate}
\par We close this paper with a short comment concerning covariance and special
frames, to wit: time plays a special r\^{o}le in cosmology. This invites the
possibility that within the fluctuation that one comes upon on extrapolating
backwards in time to the space-time patch that is occupied by the inflaton field,
there may exist special temporal effects which cannot be encoded in a covariant
formalism. In particular there may arise a term in $\delta \dot n$. This would
attenuate oscillations and, for all we know, at the early stages could compete with
the so-called friction term ($=H\dot\phi$) of chaotic inflation (where $H$ = Hubble constant). Such
a term would result in dissipative effects in which oscillations of mode density would
dissipate into the soup i.e. the mass $\mu$ could develop a frequency dependent
complex part. One should not lose sight of this possibility in developing
phenomenological implications of the picture that has been sketched. 

In conjunction with this
manuscript, Fran\c{c}ois Englert has introduced me to the recent advances in string and brane
theory concerning black hole (brane) entropy in AdS spaces\cite{LABEL15}. These latter are taken to
describe space-time outside, but in the vincinity of the black hole horizon. Could it be that our
exchange mechanism is concerned precisely with the concept expressed in this vision?

This is to express warm thanks to Fran\c{c}ois Englert for his patient explanations. In addition I
have greatly benefited from conversations and advice from Serge Massar and Philippe Spindel to whom
I here express my gratitude.

\end{document}